\documentclass[prb,twocolumn,floatfix,showpacs]{revtex4}
\usepackage{amsfonts}
\usepackage{amsmath}
\usepackage{amssymb}
\usepackage{graphicx}

\providecommand{\U}[1]{\protect\rule{.1in}{.1in}}

\begin{document}

\title[ ]{Domain Wall Dynamics near a Quantum Critical Point}
\author{Shengjun Yuan}
\affiliation{Department of Applied Physics, Zernike Institute for Advanced Materials,
University of Groningen, Nijenborgh 4, NL-9747 AG Groningen, The Netherlands}
\author{Hans De Raedt}
\affiliation{Department of Applied Physics, Zernike Institute for Advanced Materials,
University of Groningen, Nijenborgh 4, NL-9747 AG Groningen, The Netherlands}
\author{Seiji Miyashita}
\affiliation{Department of Physics, Graduate School of Science, University of Tokyo,
Bunkyo-ku,Tokyo 113-0033, Japan}
\affiliation{CREST, JST, 4-1-8 Honcho Kawaguchi, Saitama, Japan}

\begin{abstract}
We study the real-time domain-wall dynamics near a quantum critical point of
the one-dimensional anisotropic ferromagnetic spin $1/2$ chain. By numerical
simulation, we find the domain wall is dynamically stable in the
Heisenberg-Ising model. Near the quantum critical point, the width of the
domain wall diverges as $\left( \Delta -1\right) ^{-1/2}$.
\end{abstract}

\volumeyear{year}
\volumenumber{number}
\issuenumber{number}
\eid{identifier}
\date{\today }
\maketitle








\section{Introduction}

Recent progress in synthesizing materials that contain ferromagnetic chains%
\cite{Kaji05,Mito05,Kage97,Maig00} provides new opportunities to study the
quantum dynamics of atomic-size domain walls (DW). On the atomic level, a DW
is a structure that is stable with respect to (quantum) fluctuations,
separating two regions with opposite magnetization. Such a structure was
observed in the one-dimensional CoCl$_{2}\cdot2$H$_{2}$O chain\cite%
{Torrance69,Nicoli74}.

In an earlier paper\cite{Yuan06}, we studied the propagation of spin waves
in ferromagnetic quantum spin chains that support DWs. We demonstrated that
DWs are very stable against perturbations, and that the longitudinal
component of the spin wave speeds up when it passes through a DW while the
transverse component is almost completely reflected.

In this paper, we focus on the dynamic stability of the DW in the
Heisenberg-Ising ferromagnetic chain. It is known that the ground state of
this model in the subspace of total magnetization zero supports DW structures%
\cite{Gochev77,Gochev83}. However, if we let the system evolve in time from
an initial state with a DW structure and this initial state is not an
eigenstate, it must contain some excited states. Therefore, the question
whether the DW structure will survive in the stationary (long-time) regime
is nontrivial.

The question how the DW structure dynamically survives in the stationary
(long-time) region is an interesting problem. In particular, we focus on the
stability of the DW with respect to the dynamical (quantum) fluctuations as
we approach the quantum critical point (from Heisenberg-Ising like to
Heisenberg). We show that the critical quantum dynamics of DWs can be
described well in terms of conventional power laws. The behavior of quantum
systems at or near a quantum critical point is of contemporary interest\cite%
{Sachdev99}. We also show that the DW profiles rapidly become very stable as
we move away from the quantum critical point.

\section{Model}

The Hamiltonian of the system is given by\cite%
{MIKE91,Cloizeaux66,Gochev77,Gochev83,MATT81} 
\begin{equation}
H=-J\sum_{n=1}^{N-1}(S_{n}^{x}S_{n+1}^{x}+S_{n}^{y}S_{n+1}^{y}+\Delta
S_{n}^{z}S_{n+1}^{z}),  \label{H1}
\end{equation}
where $N$ indicates the total number of spins in the spin chain, and the
exchange integrals $J$ and $J\Delta$ determine the strength of the
interaction between the $x$, $y$ and $z$ components of spin $1/2$ operators $%
\mathbf{S} _{n}=\left( S_{n}^{x},S_{n}^{y}\,,S_{n}^{z}\right) $. Here we
only consider the system with the ferromagnetic ($J>0$) nearest exchange
interaction. It is well known that $\left\vert \Delta\right\vert =1$ is a
quantum critical point of the Hamiltonian in Eq. (\ref{H1}), that is, the
analytical expressions of the ground state energy for $1<\Delta$ and $%
-1<\Delta<1$ are different and singular at the points $\Delta=\pm1$\cite%
{Cloizeaux66}.

In Ref.\cite{Gochev77,Gochev83} Gochev constructed a stable state with DW
structure in both the classical and quantum treatments of the Hamiltonian (%
\ref{H1}). In the classical treatment, Gochev replaces the spin operators in
Eq. (\ref{H1}) by classical vectors of length $s$ 
\begin{equation*}
S_{n}^{z}=s\cos \theta _{n},S_{n}^{x}=s\sin \theta _{n}\cos \varphi
_{n},S_{n}^{y}=s\sin \theta _{n}\sin \varphi _{n},
\end{equation*}%
and then uses the conditions $\delta E/\delta \theta =0$ and $\varphi
_{n}=const.$ to find the ground state. In the ground state, the
magnetization per site is given by\cite{Gochev83} 
\begin{equation}
\begin{array}{c}
S_{n}^{z}=s\tanh (n-n_{0})\sigma , \\ 
S_{n}^{x}=s\cos \varphi \text{ }\mathrm{sech}(n-n_{0})\sigma , \\ 
S_{n}^{y}=s\sin \varphi \text{ }\mathrm{sech}(n-n_{0})\sigma ,%
\end{array}
\label{SzSx}
\end{equation}%
where 
\begin{equation}
\sigma =\ln [\Delta +\sqrt{\Delta ^{2}-1}],  \label{sigma1}
\end{equation}%
$\varphi $ is an arbitrary constant, and $n_{0}$ is a constant fixing the
position of the DW. The corresponding energy is 
\begin{equation}
E_{DW}=2s^{2}J\Delta \tanh \sigma .  \label{Edw}
\end{equation}%
In the quantum mechanical treatment, Gochev first constructs the
eigenfunction of a bound state of $k$ magnons\cite{Gochev83} 
\begin{equation}
\left\vert \psi _{k}\right\rangle
=A_{n}\sum_{\{m_{l}%
\}}B_{m_{1}m_{2}...m_{k}}S_{m_{1}}^{-}S_{m_{2}}^{-}...S_{m_{k}}^{-}\left%
\vert 0\right\rangle ,  \label{magnon}
\end{equation}%
where 
\begin{equation}
B_{m_{1}m_{2}...m_{k}}=\prod\limits_{i=1}^{k}v_{i}^{m_{i}},m_{i}<m_{i+1},
\label{Bmi}
\end{equation}%
\begin{equation}
v_{i}=\cosh (i-1)\sigma /\cosh (i\sigma ),  \label{vii}
\end{equation}%
\begin{equation}
A^{-2}=\prod\limits_{i=1}^{k}v_{i}^{2i}/(1-v_{i}^{2}),
\end{equation}%
and the corresponding energy is given by\cite{Gochev83} 
\begin{equation}
\epsilon _{k}=\frac{1}{2}J\Delta \tanh \sigma \tanh k\sigma .  \label{ek}
\end{equation}%
Then he demonstrated that for the infinite chain, the linear superposition 
\begin{equation}
\left\vert \phi _{n_{0}}\right\rangle =A\sum_{i=-\infty }^{\infty }\exp
\left\{ -\frac{1}{2}\sigma \left[ i+\left( \frac{1}{2}-\alpha \right) \right]
\right\} \left\vert \psi _{N_{0}+i}\right\rangle ,  \label{magnon2}
\end{equation}%
where 
\begin{equation}
n_{0}=N_{0}+\alpha ,\left\vert \alpha \right\vert \leq 1/2,N_{0}\rightarrow
\infty ,
\end{equation}%
\begin{equation}
A^{-2}=\sum_{i=-\infty }^{\infty }\exp \left\{ -\frac{1}{2}\sigma \left[
i+\left( \frac{1}{2}-\alpha \right) \right] ^{2}\right\} ,
\end{equation}%
is the quantum analog of the classical domain wall, in which $\left\langle
S_{n}^{z}\right\rangle ,\left\langle S_{n}^{x}\right\rangle ,\left\langle
S_{n}^{y}\right\rangle $ are given in Eq. (\ref{SzSx}), and the energy
coincides with Eq. (\ref{Edw}).

Gochev's work confirmed the existence of the DW structures in the
one-dimensional ferromagnetic quantum spin $1/2$ chain. In the infinite
chain, the exact quantum analog of classical DW is represented by $%
\left\vert \phi_{n_{0}}\right\rangle $. In the finite chain, the DW
structure exists as a bound $k$-magnon state $\left\vert
\psi_{k}\right\rangle $. The main difference between these two states is the
distribution of magnetization in the XY plane. In the infinite chain, the
change of the magnetization occurs in three dimensions, according to Eq. (%
\ref{SzSx}), but in the finite chain $\left\langle S_{n}^{x}\right\rangle
=\left\langle S_{n}^{y}\right\rangle =0$ for all spins.

\begin{figure}[t]
\begin{center}
\includegraphics[
width=8cm
]{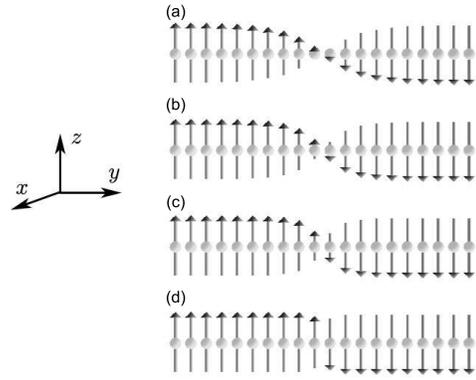}
\end{center}
\caption{The magnetization $\left\langle S_{n}^{z}\right\rangle $\ in the
ground state of the subspace of total magnetization $M=0$, generated by the
power method. The parameters are: (a) $\Delta=1.05$, (b) $\Delta=1.1$, (c) $%
\Delta=1.2$, (d) $\Delta=2$. The total number of spins in the spin chain is $%
N=20$. It is clear that there is a DW at the centre of the spin chain.
Furthermore there is no structure in the XY plane, that is, $\left\langle
S_{n}^{x}\right\rangle =\left\langle S_{n}^{y}\right\rangle =0$.}
\label{figure0}
\end{figure}

\begin{figure}[t]
\begin{center}
\includegraphics[
width=8cm
]{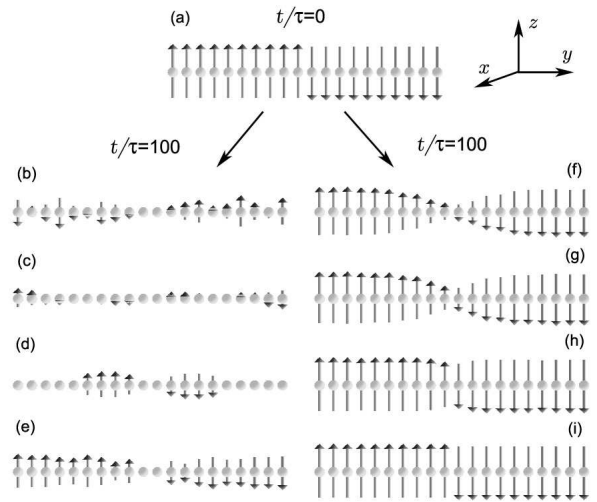}
\end{center}
\caption{Top picture (a): Initial spin configuration at time $t/\protect\tau%
=0$; Bottom pictures (b,c,d,e,f,g,h,i): Spin configuration at time $t/%
\protect\tau=100$; Bottom left pictures (b,c,d,e): DW structures disappear
or are not stable. The parameters are: (b) $\Delta=0$ (XY model), (c) $%
\Delta=0.5$ (Heisenberg-XY model), (d) $\Delta=1$ (Heisenberg model), (e) $%
\Delta=1.05$ (Heisenberg-Ising model); Bottom right pictures (f,g,h,i): DW
structures are dynamically stable in the Heisenberg-Ising model. The
parameters are: (f) $\Delta=1.1$, (g) $\Delta=1.2$, (h) $\Delta=2$, (i) $%
\Delta=20$. The total number of spins in the spin chain is $N=20$.}
\label{figure1}
\end{figure}

Now we consider $\left\langle S_{n}^{z}\right\rangle $ of the bound state $%
\left\vert \psi_{k}\right\rangle $ in the case that the number of flipped
spin is half of the total spins, i.e., $k=N/2$ and $N$ is an even number.
Even though the formal expression for $\left\vert \psi_{k}\right\rangle $ is
known, the expression for $\left\langle S_{n}^{z}\right\rangle $ in this
state (for finite and infinite chains) is not known. For finite $N$, the
ground state in the subspace of total magnetization $M=0$ can, in principle,
be calculated from Eq. (\ref{magnon}). However, this requires a numerical
procedure and we loose the attractive features of the analytical approach.
Indeed, it is more efficient to use a numerical method and compute directly
the ground state in the subspace of total magnetization $M=0$. In Fig. \ref%
{figure0}, we show some representative results as obtained by the power
method\cite{Wilkinson99} for a chain of $N=20$ spins. In all cases, the
domain wall is well-defined. Obviously, because we are considering the
system in the ground state, the magnetization profile will not change during
the time evolution.

To inject a DW in the spin chain, we take
the state $\vert \Phi \rangle $ with the left half of the spins up and the other half down
as the initial state (see Fig. \ref{figure1}(a) for $N=20$).
The state $\vert \Phi \rangle $ corresponds to the
state with the largest weight in the bound state
$\vert \psi _{k}\rangle $ with $k=N/2$, because $%
\vert B_{m_{1}m_{2}...m_{k}}\vert ^{2}$ reaches the maximum if $%
m_{i}=i$ for all $i=1,2,..,N/2$ (note $\vert v_{i}\vert <1$). It
is clear that $\vert \Phi \rangle $ is not an eigenstate of the
Hamiltonian in Eq. (\ref{H1}). The energy of $\vert \Phi
\rangle $, relative to the ferromagnetic ground state, is $J\Delta /2$, and its
spread $( \langle \Phi \vert H^{2}\vert \Phi
\rangle -\langle \Phi \vert H\vert \Phi \rangle
^{2}) ^{1/2}=J/2$.
In Table \ref{tablea}, we list some representative values of the energy
in the initial state (see Fig. \ref{figure1}(a))
and
in the ground state of subspace $M=0$ (see Fig. \ref{figure0}).

A priori, there is no reason why the DW
of the initial state $\vert \Phi \rangle $
should relax to a DW profile that is dynamically stable.
For $\Delta \simeq 1$, the difference between
energy of the initial state and the ground state energies
for $N=16,18,20,22$ is relatively large and
the relative spread in energy ($1/\Delta$) is large also,
suggesting that near the quantum critical point, the initial state
may contain a significant amount of excited states.
Therefore, it is not evident that a DW structure will survive in the long-time regime.
In fact, from the numbers in Table I, one cannot predict whether or not the DW will be stable.
For instance, for $\Delta=1.05$ and $N=16,18,20$, the DW is not dynamically stable
whereas for $N=22$ it is stable but the energies (see first line in Table I)
do not give a clue as to why this should be the case.
On the other hand, by solving the time dependent Schr\"{o}dinger equation (TDSE),
it is easy to see if the DW is dynamically stable or not.

\begin{table}[t]
\caption{The energy $E=J\Delta /2$ of the initial state $\vert \Phi\rangle$ (see Fig. \protect\ref{figure1}(a))
and the ground state $E_g^{(N)}$ in the $M=0$ subspace, both
relative to the ground state energy of the ferromagnet.
}
\label{tablea}
\begin{center}
\begin{ruledtabular}
\begin{tabular}{cccccc}
$\Delta$ & $E$ & $E_g^{(16)}$& $E_g^{(18)}$& $E_g^{(20)}$& $E_g^{(22)}$\\
\hline
%
$1.05$ & $0.53$     & $0.16$ & $0.16$ & $0.16$ & $0.16$ \\
$1.1$  &  $0.55$    & $0.23$ & $0.23$ & $0.23$ & $0.23$ \\
$2$    &     $1.00$ & $0.87$ & $0.87$ & $0.87$ & $0.87$ \\
$5$    &   $2.50$   & $2.45$ & $2.45$ & $2.45$ & $2.45$
\end{tabular}
\end{ruledtabular}
\end{center}
\end{table}


\section{Dynamically Stable Domain Walls}

We solve the TDSE of the whole system with the Hamiltonian in Eq. (\ref{H1})
and study the time-evolution of the magnetization at each lattice site. The
numerical solution of the TDSE is performed by the Chebyshev polynomial
algorithm, which is known to yield extremely accurate independent of the
time step used\cite{TALE84,LEFO91,IITA97,DOBR03}. We adopt open boundary
conditions, not periodic boundary conditions, because the periodic boundary
condition would introduce two DWs in the initial state. In this paper, we
display the results at time intervals of $\tau =\pi /5J$, and use units such
that $\hbar =1$ and $J=1$.

The initial state of the system is shown in Fig. \ref{figure1}(a). The spins
in the left part ($n=1$ to $10$) of the spin chain are all "spin-up" and the
rest ($n=11$ to $20$) are all "spin-down". Here "spin-up" or "spin-down"
correspond to the eigenstates of the single spin $1/2$ operator $S_{n}^{z}$.

Whether the DW at the centre of the spin chain is stable or unstable depends
on the value of $\Delta$. In Fig. \ref{figure1}(b,c,d,e,f,g,h, and i), we
show the states of the system as obtained by letting the system evolve over
a fairly long time ($t=500J/\pi$). It is clear that the DW totally
disappears for $0\leq\Delta\leq1$, that is, in the XY, Heisenberg-XY and
Heisenberg spin $1/2$ chain, the DW structures are not stable. For the
Heisenberg-Ising model ($\Delta>1$), the DW remains stable when $%
t\geq500J/\pi$ (see Ref.\cite{Yuan06}), and its structure is more sharp and
clear if $\Delta$ is larger, so we will concentrate on the cases $\Delta>1$.
One may note that the values of $\Delta$ in Fig. \ref{figure1}(e,f,g,h) are
the same as in Fig. \ref{figure0}(a,b,c,d), but that the distributions of
the magnetization are similar but not the same. This is because the energy
is conserved during the time evolution and the system, which starts from the
initial state shown in Fig. \ref{figure1}(a), will never relax to the ground
state of the subspace with the total magnetization $M=0$.

\begin{figure}[t]
\begin{center}
\includegraphics[
width=8cm
]{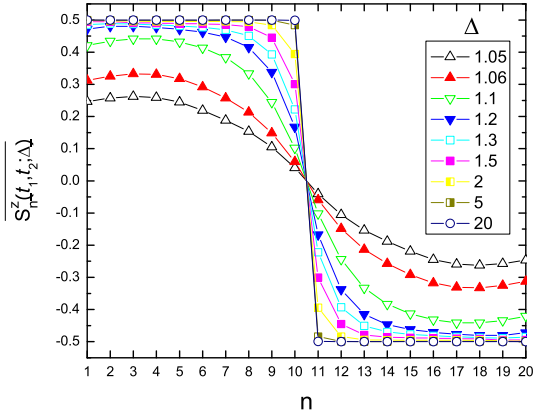}
\end{center}
\caption{(Color online) $\overline{S_{n}^{z}\left( t_{1},t_{2};\Delta
\right) }$ as a function of $n$ for different $\Delta $. Here $t_{1}=101%
\protect\tau $, $t_{2}=200\protect\tau $. We show the data for $\Delta
=1.05,1.06,1.1,1.2,1.3,1.5,2,5$ and $20$ only. The total number of spins in
the spin chain is $N=20$.}
\label{figure2}
\end{figure}

\begin{figure}[t]
\begin{center}
\includegraphics[
width=8cm
]{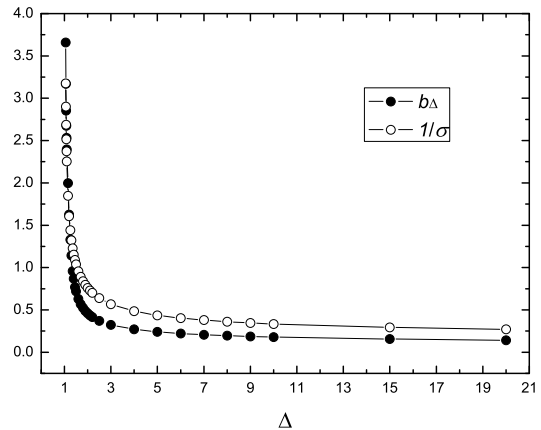}
\end{center}
\caption{Comparison of $b_{\Delta}$ and $1/\protect\sigma$ as a function of $%
\Delta$. The total number of spins in the spin chain is $N=20$.}
\label{figsigma}
\end{figure}

In order to get a quantitative expression of the width of DW, we first
introduce the quantity $\overline{S_{n}^{z}\left( t_{1},t_{2};\Delta \right) }$ (%
$n=1,2,...,N$) as the time average of the expectation value $\left\langle
S_{n}^{z}\left( t\right) \right\rangle $ of $n$th spin: 
\begin{equation}
\overline{S_{n}^{z}\left( t_{1},t_{2};\Delta \right) }\equiv \frac{%
\int_{t_{1}}^{t_{2}}\left\langle S_{n}^{z}\left( t\right) \right\rangle dt}{%
t_{2}-t_{1}}.  \label{Sz1}
\end{equation}%
We take the average in Eq. (\ref{Sz1}) over a long period during which the
DW is dynamically stable. In Fig. \ref{figure2}, 
we show some results of $\overline{S_{n}^{z}\left( t_{1},t_{2};\Delta
\right) }$ for the Heisenberg-Ising model, where we take $t_{1}=101\tau $, $%
t_{2}=200\tau $ and various $\Delta $. We find that each curve in Fig. \ref%
{figure2} is symmetric about the line $n=\left( N+1\right) /2$, and can be
fitted well by the function 
\begin{equation}
\overline{S_{n}^{z}\left( t_{1},t_{2};\Delta \right) }= a_{\Delta
}\tanh \left[ \frac{n-(N+1)/2}{b_{\Delta }}\right] .  \label{sznfit}
\end{equation}%
The values of $\Delta $ we used and the corresponding values of $a_{\Delta }$%
, $b_{\Delta }$ are shown in Table \ref{table1}. As we mentioned earlier,
Gochev\cite{Gochev83} constructed an eigenstate of the one-dimensional
anisotropic ferromagnetic spin $1/2$ chain in which the mean values $%
S_{n}^{z}$, $S_{n}^{x}$ and $S_{n}^{y}$ coincide with the stable DW
structure in the classical spin chain, that is 
\begin{equation}
\left\langle S_{n}^{z}\right\rangle =\frac{1}{2}\tanh (n-n_{0})\sigma ,
\label{Szn2}
\end{equation}%
where $n_{0}$ is the position of the DW (in our notation, this is $\left(
N+1\right) /2$). The fitted form of $\overline{S_{n}^{z}\left(
t_{1},t_{2};\Delta \right) }$ in Eq. (\ref{sznfit}) is similar to Eq. (\ref%
{Szn2}). From Table \ref{table1}, it is clear that as $\Delta $ increases, $%
\left\vert a_{\Delta }\right\vert $ converges to $1/2$, in agreement with
Eq. (\ref{Szn2}). From the comparison of $b_{\Delta }$ and $1/\sigma $ in
Fig. \ref{figsigma}, it is clear that the dependence on $\Delta $ is
qualitatively similar but not the same. This is due to the fact that
Gochev's solution is for a DW in the ground state whereas we obtain the DW
by relaxation of the state shown in Fig. \ref{figure1}(a).

We want to emphasize that the meaning of $\overline{S_{n}^{z}\left(
t_{1},t_{2};\Delta\right) }$ in Eq. (\ref{sznfit}) is different from $%
\left\langle S_{n}^{z}\right\rangle $ in Eq. (\ref{Szn2}). The former
describes the mean value of $\left\langle S_{n}^{z}\left( t\right)
\right\rangle $ in a state with dynamical fluctuations, while the latter
describes the distribution of $\left\langle S_{n}^{z}\right\rangle $ in an
exact eigenstate without dynamical fluctuations.

\begin{table}[t]
\caption{The values of $\Delta$ we used in our simulations and the
corresponding $a_{\Delta}$, $b_{\Delta}$ fitted by Eq. (\protect\ref{sznfit}%
) for a spin chain of $N=20$ spins.}
\label{table1}
\begin{center}
\begin{ruledtabular}
\begin{tabular}{ccc|ccc}%
$\Delta$ & ${a}_{\Delta}$ & ${b}_{\Delta}$ & $\Delta$ & ${a}_{\Delta}$ &
${b}_{\Delta}$\\\hline
$1.05$ & $-0.263$ & $3.659$ & $1.8$ & $-0.493$ & $0.524$\\
$1.06$ & $-0.330$ & $3.171$ & $1.9$ & $-0.494$ & $0.488$\\
$1.07$ & $-0.377$ & $2.850$ & $2$ & $-0.495$ & $0.460$\\
$1.08$ & $-0.406$ & $2.673$ & $2.1$ & $-0.495$ & $0.436$\\
$1.09$ & $-0.424$ & $2.534$ & $2.2$ & $-0.496$ & $0.416$\\
$1.1$ & $-0.435$ & $2.396$ & $2.5$ & $-0.497$ & $0.370$\\
$1.15$ & $-0.462$ & $1.996$ & $3$ & $-0.498$ & $0.322$\\
$1.2$ & $-0.471$ & $1.626$ & $4$ & $-0.499$ & $0.270$\\
$1.25$ & $-0.476$ & $1.330$ & $5$ & $-0.499$ & $0.240$\\
$1.3$ & $-0.479$ & $1.142$ & $6$ & $-0.500$ & $0.220$\\
$1.35$ & $-0.481$ & $0.959$ & $7$ & $-0.500$ & $0.206$\\
$1.4$ & $-0.483$ & $0.869$ & $8$ & $-0.500$ & $0.195$\\
$1.45$ & $-0.485$ & $0.770$ & $9$ & $-0.500$ & $0.187$\\
$1.5$ & $-0.487$ & $0.719$ & $10$ & $-0.500$ & $0.179$\\
$1.6$ & $-0.489$ & $0.629$ & $15$ & $-0.500$ & $0.156$\\
$1.7$ & $-0.491$ & $0.568$ & $20$ & $-0.500$ & $0.141$

\end{tabular}
\end{ruledtabular}
\end{center}
\end{table}

\begin{figure}[t]
\begin{center}
\includegraphics[
width=8cm
]{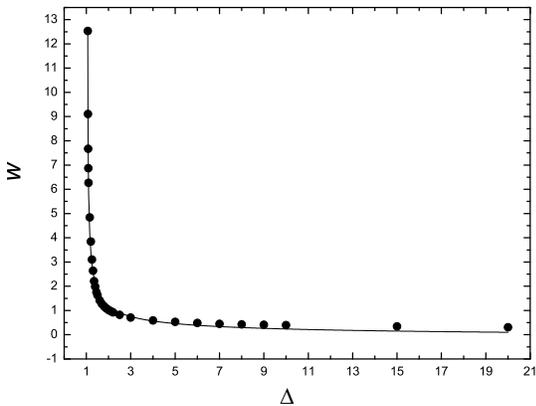}
\end{center}
\caption{The DW\ width as a function of $\Delta $ in a spin chain of $N=20$
spins. The black dots are the simulation data and the solid line is given by 
$W\left( \Delta \right) =A_{N}/\ln \left\{ \Delta -\protect\epsilon _{N}+%
\left[ \left( \Delta -\protect\epsilon _{N}\right) ^{2}-1\right]
^{1/2}\right\} +B_{N}$ with $\protect\epsilon _{N}=0.046\pm 0.001,$ $%
A_{N}=2.16\pm 0.06$ and $B_{N}=-0.485\pm 0.068$. }
\label{figure3}
\end{figure}

\begin{figure*}[t]
\begin{center}
\mbox{
\includegraphics[width=8cm]{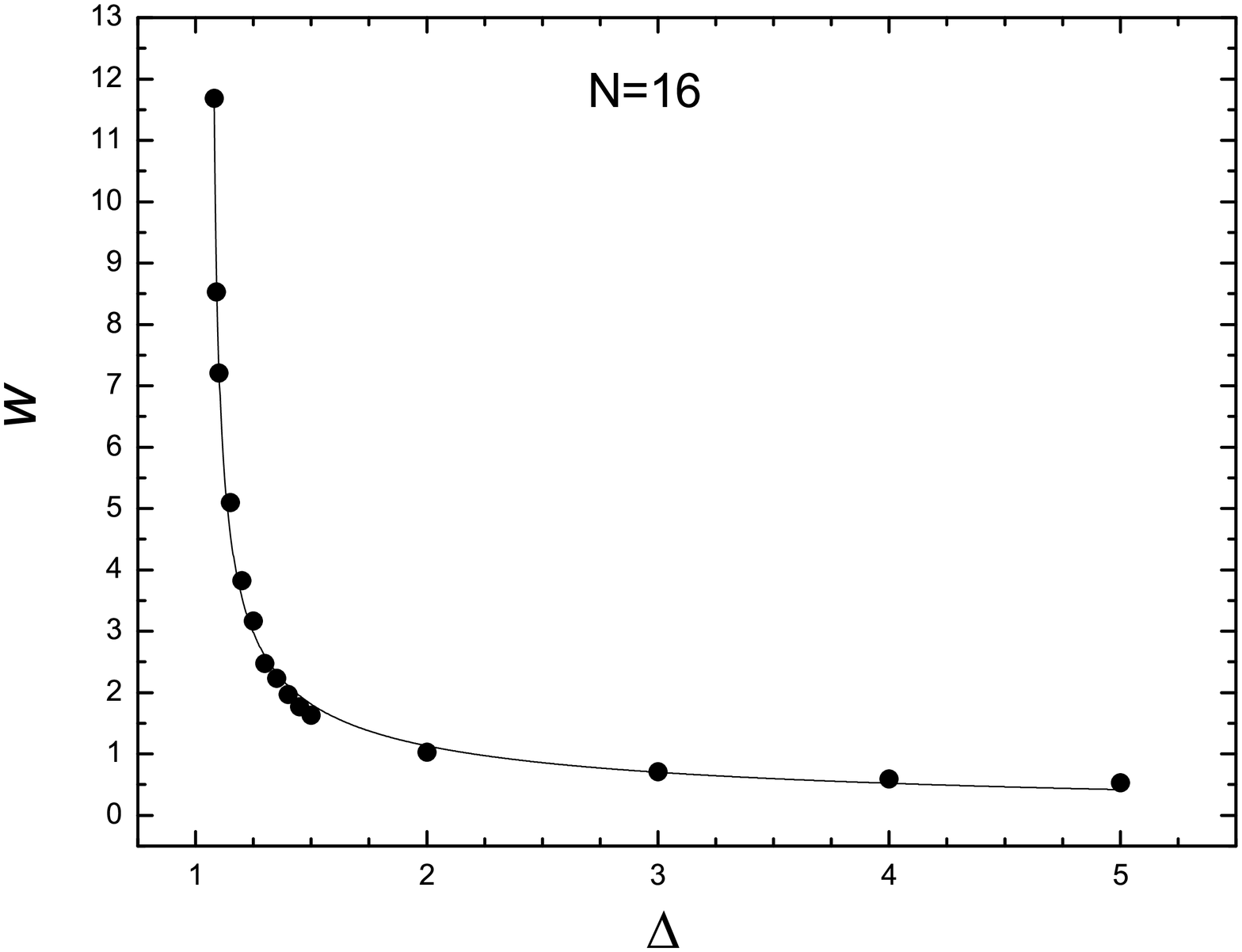}
\includegraphics[width=8cm]{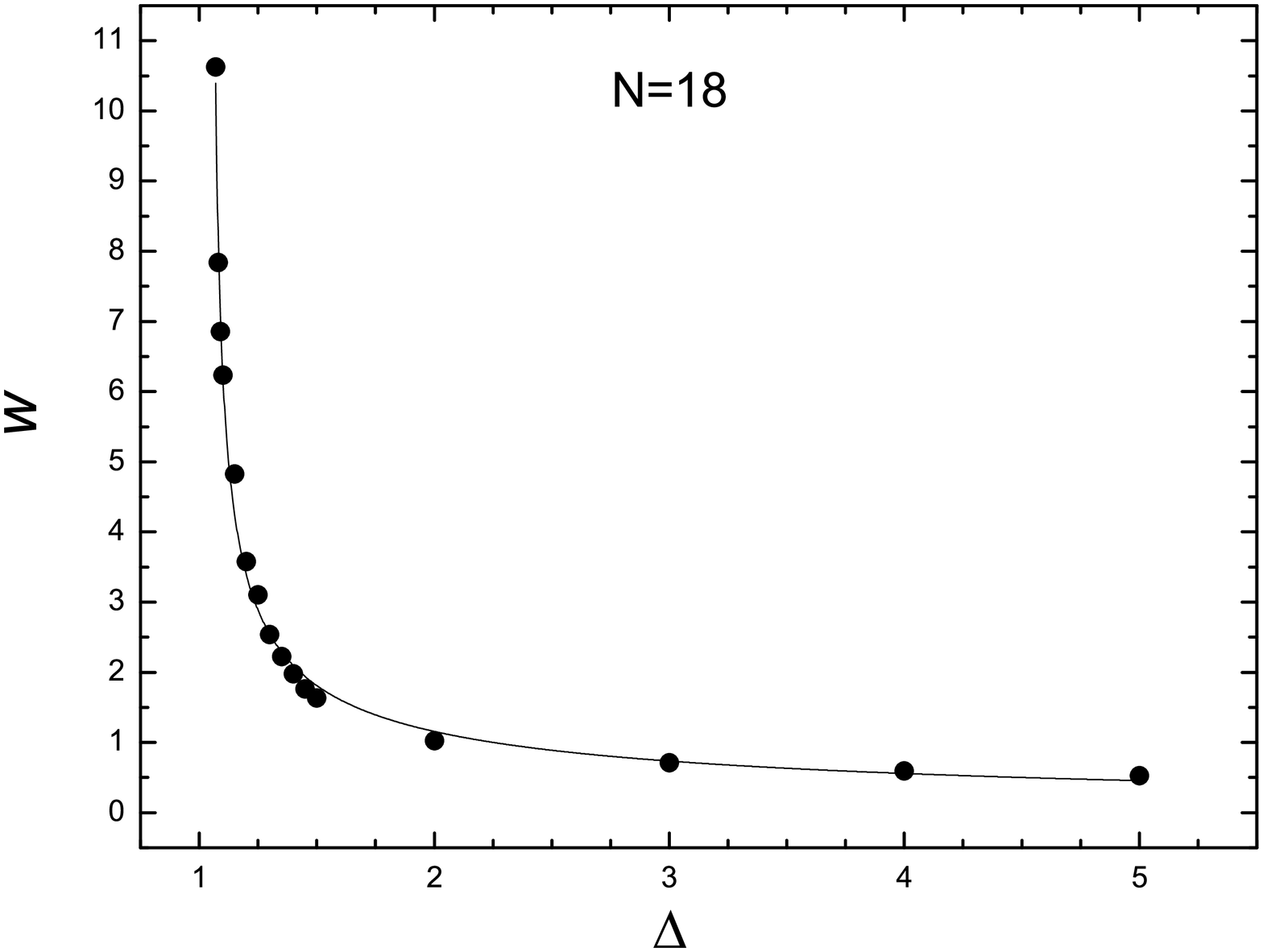}
} 
\mbox{
\includegraphics[width=8cm]{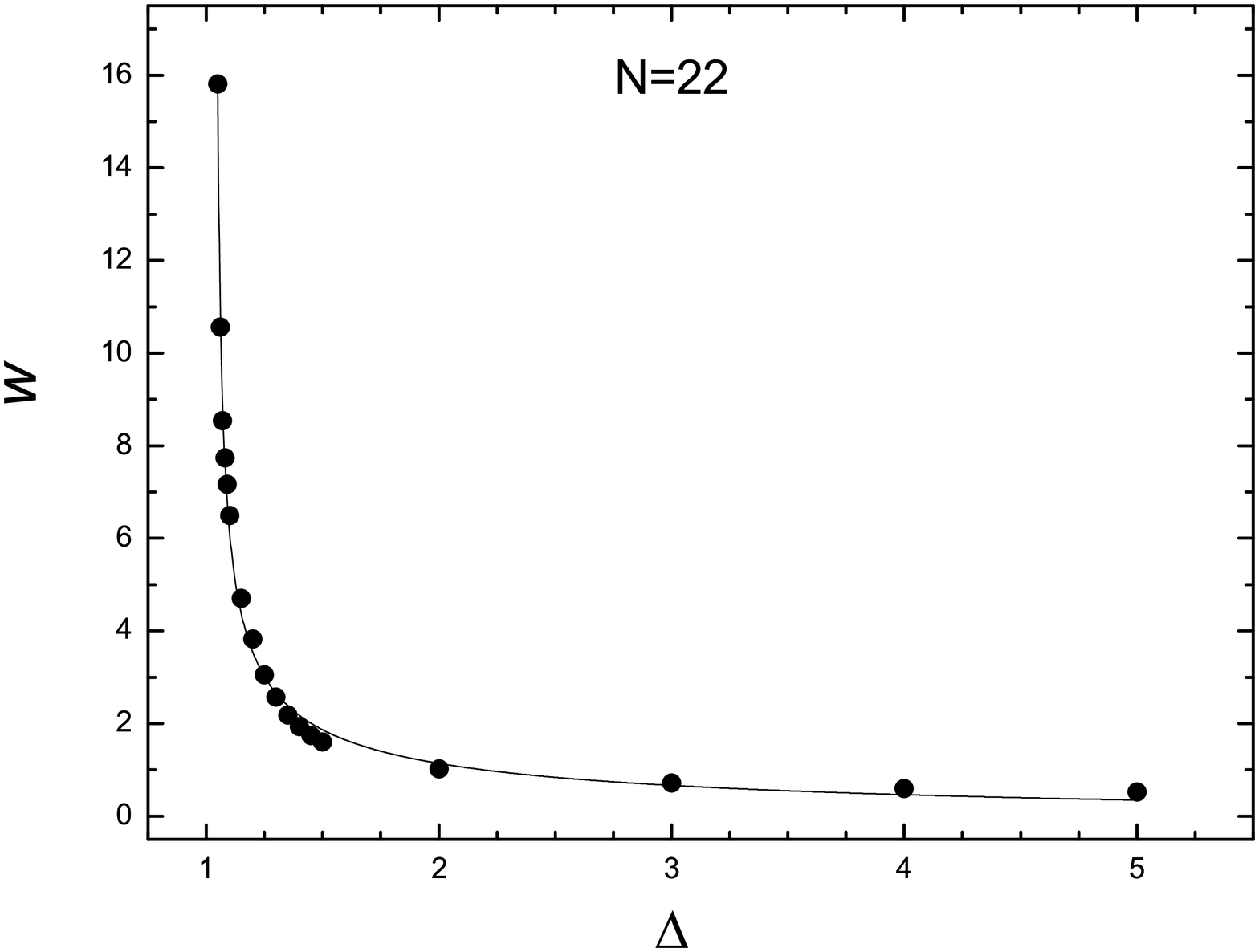}
\includegraphics[width=8cm]{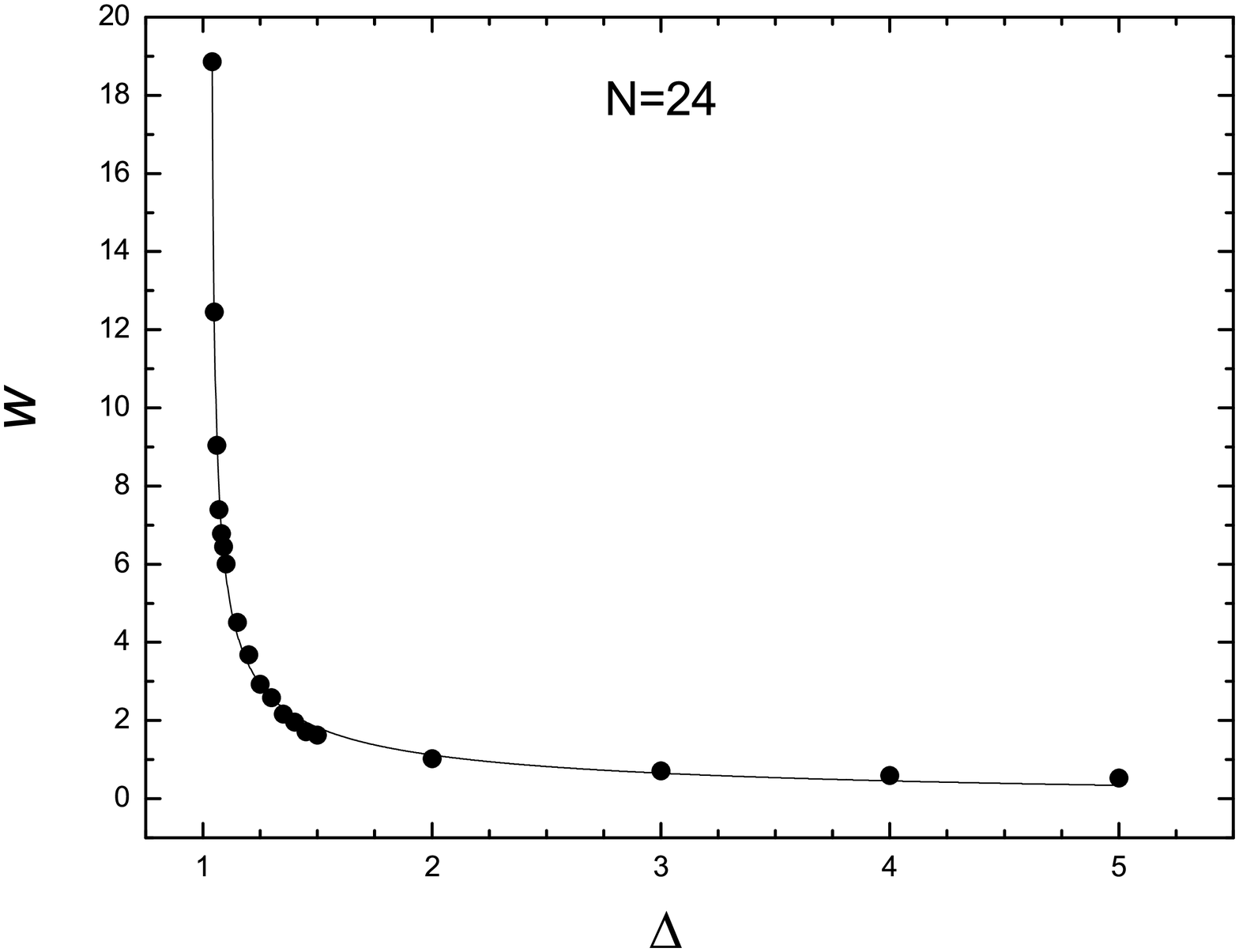}
}
\end{center}
\caption{The DW\ width as a function of $\Delta $ in a spin chain of $N=16$, 
$18$, $22$, and $24$ spins. The black dots are the simulation data and the
solid line in each panel is given by Eq. (\protect\ref{W1}).}
\label{fitdifn}
\end{figure*}

\begin{table}[t]
\caption{The values of $\protect\epsilon _{N}$, $A_{N}$ and $B_{N}$ in Eq. (%
\protect\ref{W1}) for a spin chain of $N=16$, $18$, $20$, $22$ and $24$
spins. For the fits, we used all the data for $\Delta \leq 5$.}
\label{table2}
\begin{center}
\begin{ruledtabular}
\begin{tabular}{cccc}
$N$ & $\epsilon _{N}$ & $A_{N}$ & $B_{N}$\\\hline
$16$ & $0.065\pm0.001$ & $2.08\pm0.10$ & $-0.493\pm0.142$\\
$18$ & $0.052\pm0.002$ & $2.07\pm0.11$ & $-0.450\pm0.152$\\
$20$ & $0.045\pm0.002$ & $2.22\pm0.09$ & $-0.556\pm0.133$\\
$22$ & $0.040\pm0.001$ & $2.36\pm0.08$ & $-0.689\pm0.140$\\
$24$ & $0.033\pm0.001$ & $2.34\pm0.06$ & $-0.681\pm0.127$
\end{tabular}
\end{ruledtabular}
\end{center}
\end{table}

\begin{figure}[t]
\begin{center}
\includegraphics[
width=8cm
]{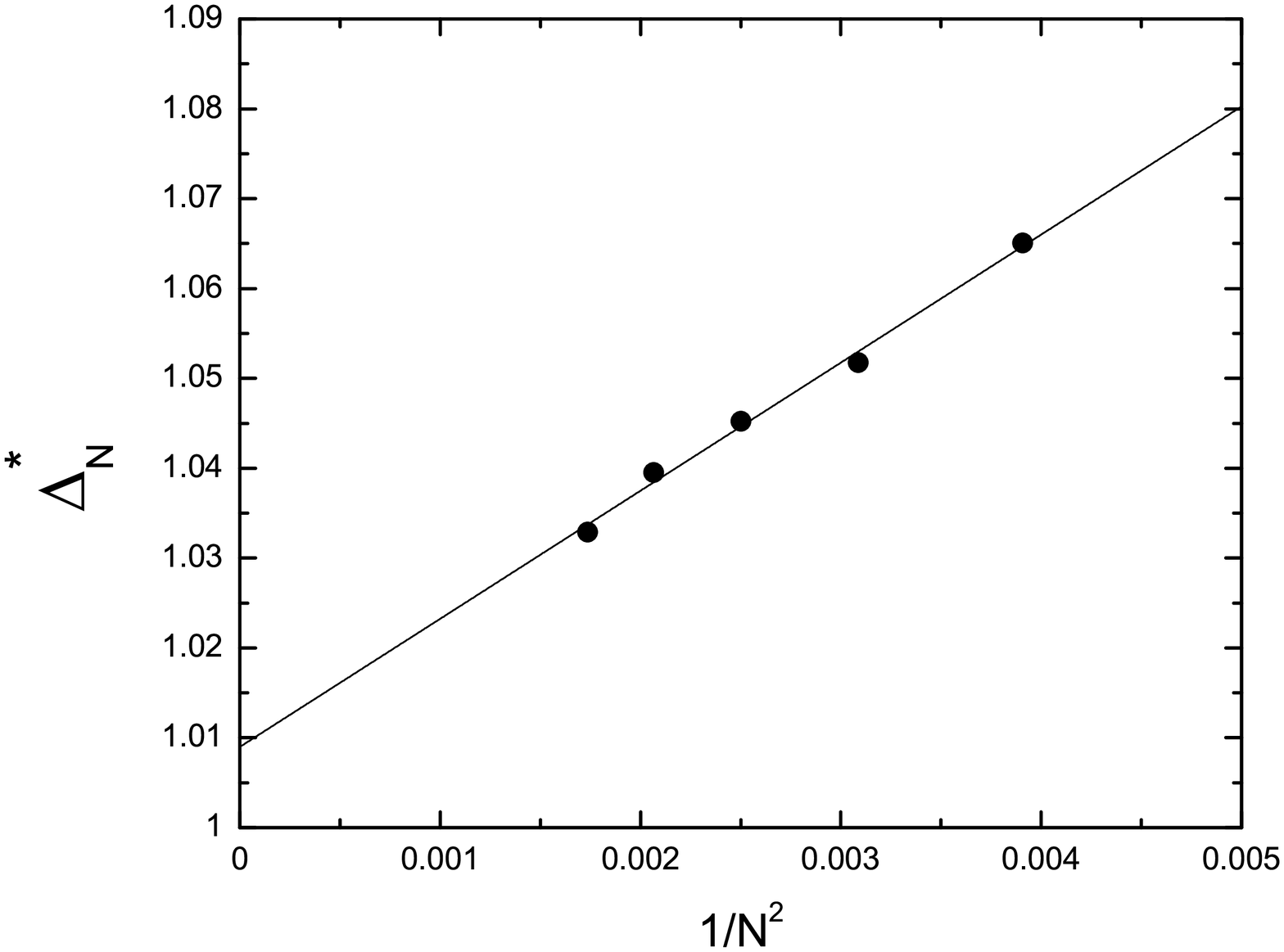}
\end{center}
\caption{Fit of $\Delta _{N}^{\ast }$ to $\Delta ^{\ast }+\protect\lambda %
\cdot N^{-2}$ with $\Delta ^{\ast }=1.009\pm 0.002$, and $\protect\lambda %
=14.253\pm 0.660$.}
\label{fitdelta0}
\end{figure}

\begin{figure}[t]
\begin{center}
\includegraphics[
width=8cm
]{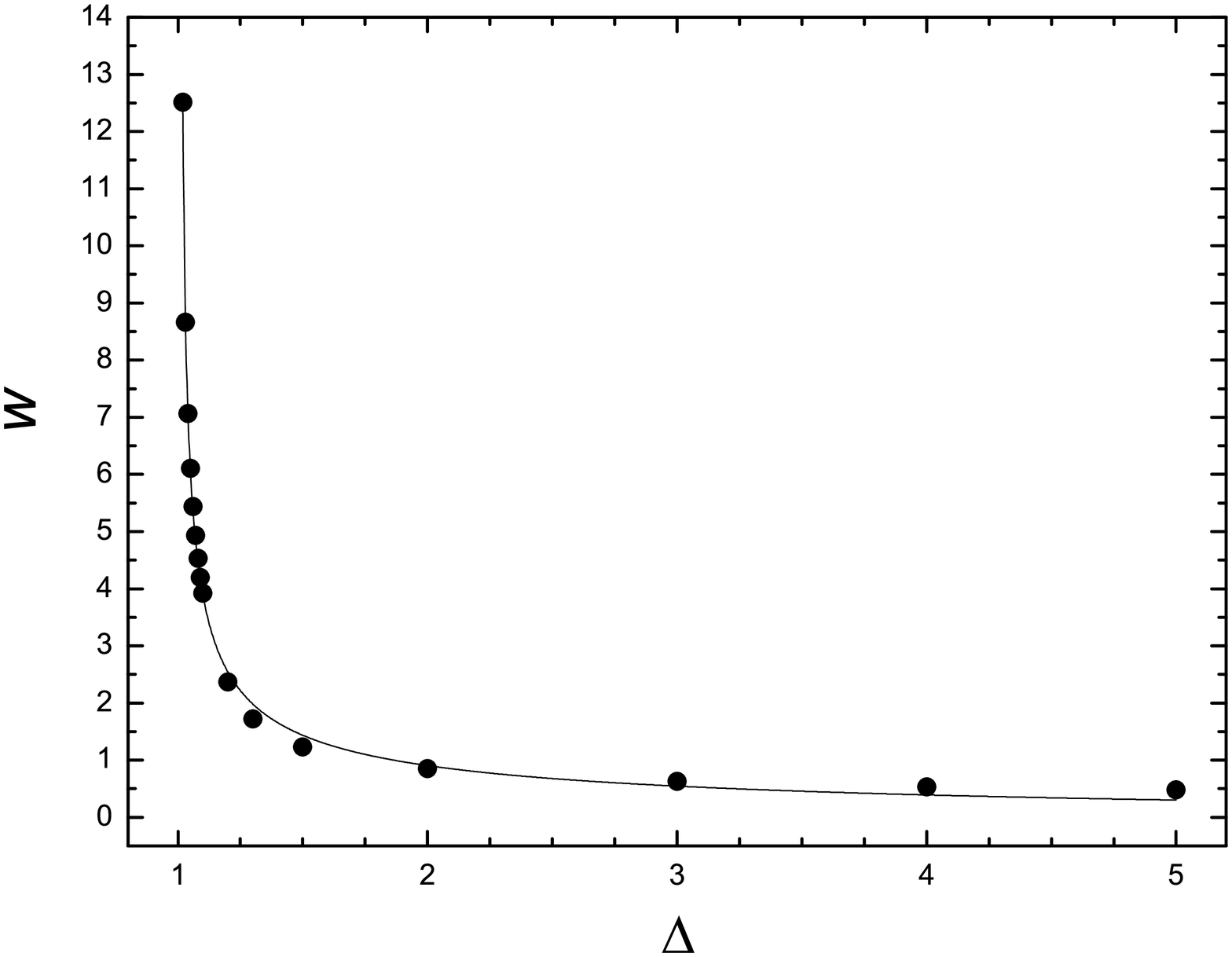}
\end{center}
\caption{The DW\ width as a function of $\Delta $ ($1.06\leq \Delta \leq 20$%
) in the ground state of subspace $M=0$ in a spin chain of $N=20$ spins. The
black dots are the simulation data and the solid line is given by Eq. (%
\protect\ref{W1}), with $\protect\epsilon _{N}=0.010\pm 0.001,$ $%
A_{N}=1.87\pm 0.04$ and $B_{N}=-0.550\pm 0.079$. }
\label{widthpm20}
\end{figure}

Next we introduce a definition of the DW width. From Eq. (\ref{sznfit}), we
can find $n_{1}$ and $n_{2}$ which satisfy 
\begin{align}
\overline{S_{n_{1}}^{z}\left( t_{1},t_{2};\Delta \right) }& =1/4,  \notag \\
\overline{S_{n_{2}}^{z}\left( t_{1},t_{2};\Delta \right) }& =-1/4,
\end{align}%
that is, when $\left\vert \overline{S_{n}^{z}\left( t_{1},t_{2};\Delta
\right) }\right\vert $ equals half of its maximum value ($1/2$). Here $n_{1}$
and $n_{2}$ are not necessarily integer numbers. Now we can define the DW
width $W$ as the distance between $n_{1}$ and $n_{2}$: 
\begin{equation}
W=\left\vert n_{1}-n_{2}\right\vert .
\end{equation}

Clearly, the width of the DW becomes ill-defined if it approaches the size
of the chain. On the other hand, the computational resources (mainly
memory), required to solve the TDSE, grow exponentially with the number of
spins in the chain. These two factors severely limit the minimum difference
between $\Delta$ and the quantum critical point ($\Delta =1$) that yields meaningful
results for the width of the DW. Indeed, for fixed $N$, $\Delta $ has to be
larger than the "effective" critical value for the finite system in order
for the DW width to be smaller than the system size. Although the system
sizes that are amenable to numerical simulation are rather small for
present-day "classical statistical mechanics" standards, it is nevertheless
possible to extract from these simulations useful information about the
quantum critical behavior of the dynamically stable DW.

In Fig. \ref{figure3}, we plot $W$ as a function of $\Delta $ ($1.06\leq
\Delta \leq 20$). By trial and error, we find that all the data can be
fitted very well by the function 
\begin{equation}
W\left( \Delta \right) =\frac{A_{N}}{\ln \left\{ \Delta -\epsilon _{N}+\left[
\left( \Delta -\epsilon _{N}\right) ^{2}-1\right] ^{1/2}\right\} }+B_{N},
\label{W1}
\end{equation}%
where $\epsilon _{N}$, $A_{N}$ and $B_{N}$ are fitting parameters. As shown
in Fig. \ref{fitdifn}, all the data for $N=16$, $18$, $22$, $24$ and $\Delta
\leq 5$ fit very well to Eq. (\ref{W1}). The results of these fits are
collected in Table \ref{table2}.

To analyze the finite-size dependence in more detail, we adopt the standard
finite-size scaling hypothesis\cite{Landau2000}. We assume that in the
infinite system, the DW width plays the role of the correlation length, that
is, we assume that%
\begin{equation}
W\left( \Delta \right) \sim W_{0}(\Delta -1)^{-\nu },  \label{W2}
\end{equation}%
where $\nu$ is a critical exponent.
Finite-size scaling predicts that the effective critical value
$\Delta _{N}^{\ast }=1+\epsilon _{N}$ where $\epsilon _{N}$
is proportional to $N^{-1/\nu}$.
Taking $\nu=1/2$, Fig. \ref{fitdelta0} shows that
$\Delta _{N}^{\ast }$ converges to one as $N$ increases.

As a check on the fitting procedure, we apply it to the data obtained by
solving for the ground state in the $M=0$ subspace. In view of Eq. (\ref{Sz1})
and (\ref{sznfit}), we may expect that Eq. (\ref{W1}) fits the data very
well and, as shown in Fig. \ref{widthpm20}, this is indeed the case.

\begin{figure}[t]
\begin{center}
\includegraphics[width=8cm]{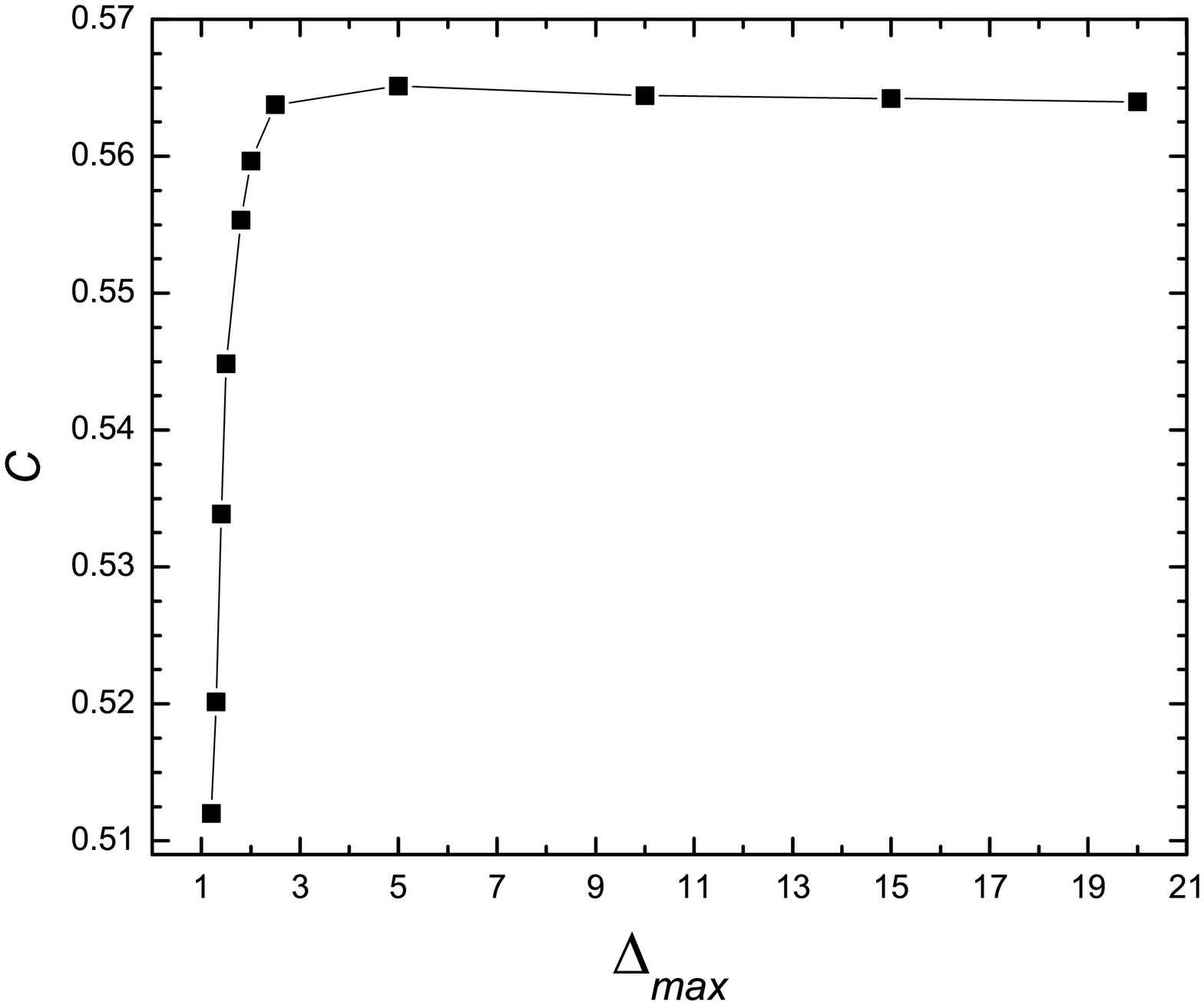}
\end{center}
\caption{The exponent $C$ as a function of $\Delta _{\max }$ in a spin chain
of $N=20$ spins. The exponent $C$ obtained by fitting the DW width to
Eq.~(\protect\ref{W3}), with $\Delta _{N=20}^{\ast }$
as obtained from the fit shown in Fig.~\ref{fitdelta0},
for $\Delta$ in the range $\left[ 1.06,\Delta _{\max }\right] $. }
\label{exponent}
\end{figure}

If we fit the data to
\begin{equation}
W\left( \Delta \right) =W_{0}\left( \Delta -\Delta _{N}^{\ast }\right) ^{-C}.
\label{W3}
\end{equation}%
without assuming a priori value $C$,
we find that $C$ depends on the range of $\Delta$ that was used
in the fit, as shown in Fig. \ref{exponent}.
Remarkably, we find that $C\approx0.57$ if we fit the data for a large range
of $\Delta$'s and that $C$ approaches $1/2$ if we restrict the value of $\Delta$
to the vicinity of the critical point.

\section{The Stability of Domain Walls}

To describe the stability of the DW structure, we introduce $%
\delta_{n}\left( \Delta\right) $ ($n=1,2,...,N$): 
\begin{equation}
\delta_{n}\left( \Delta\right) =\sqrt{\overline{\left[ S_{n}^{z}\left(
t_{1},t_{2};\Delta\right) \right] ^{2}}-\overline{S_{n}^{z}\left(
t_{1},t_{2};\Delta\right) }^{2}},  \label{delta1}
\end{equation}
where 
\begin{equation}
\overline{\left[ S_{n}^{z}\left( t_{1},t_{2};\Delta\right) \right] ^{2} }%
\equiv\frac{\int_{t_{1}}^{t_{2}}\left\langle S_{n}^{z}\left( t\right)
\right\rangle ^{2}dt}{t_{2}-t_{1}}.
\end{equation}
In order to show the physical meaning of $\delta_{n}$, we write $%
\left\langle S_{n}^{z}\left( t\right) \right\rangle $ as 
\begin{equation}
\left\langle S_{n}^{z}\left( t\right) \right\rangle \equiv C_{n}+\Omega
_{n}\left( t\right) ,
\end{equation}
where $C_{n}$ is a constant and $\Omega_{n}\left( t\right) $ is a
time-dependent term. Then Eq. (\ref{delta1}) becomes 
\begin{equation}
\delta_{n}\left( \Delta\right) =\left\{ \frac{\int_{t_{1}}^{t_{2}}
\Omega_{n}^{2}\left( t\right) dt}{t_{2}-t_{1}}-\left[ \frac{\int_{t_{1}
}^{t_{2}}\Omega_{n}\left( t\right) dt}{t_{2}-t_{1}}\right] ^{2}\right\}
^{1/2}.  \label{delta2}
\end{equation}

\begin{figure}[t]
\begin{center}
\includegraphics[
width=8cm
]{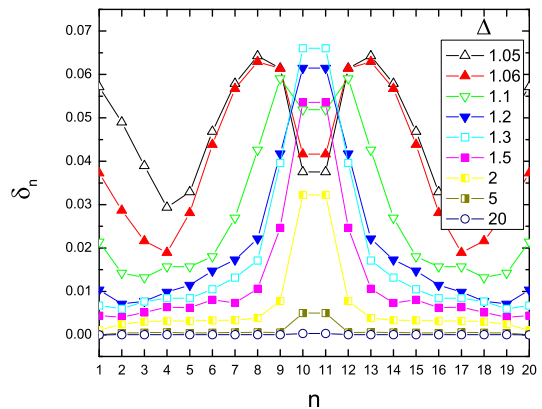}
\end{center}
\caption{(Color online) $\protect\delta_{n}\left( \Delta\right) $ as a
function of $n$ for different $\Delta$. Here $t_{1}=101\protect\tau$, $%
t_{2}=200\protect\tau$. We only show the data for $%
\Delta=1.05,1.06,1.1,1.2,1.3,1.5,2,5$ and $20$. The total number of spins in
the spin chain is $N=20$.}
\label{figure4}
\end{figure}

\begin{figure}[t]
\begin{center}
\includegraphics[
width=8cm
]{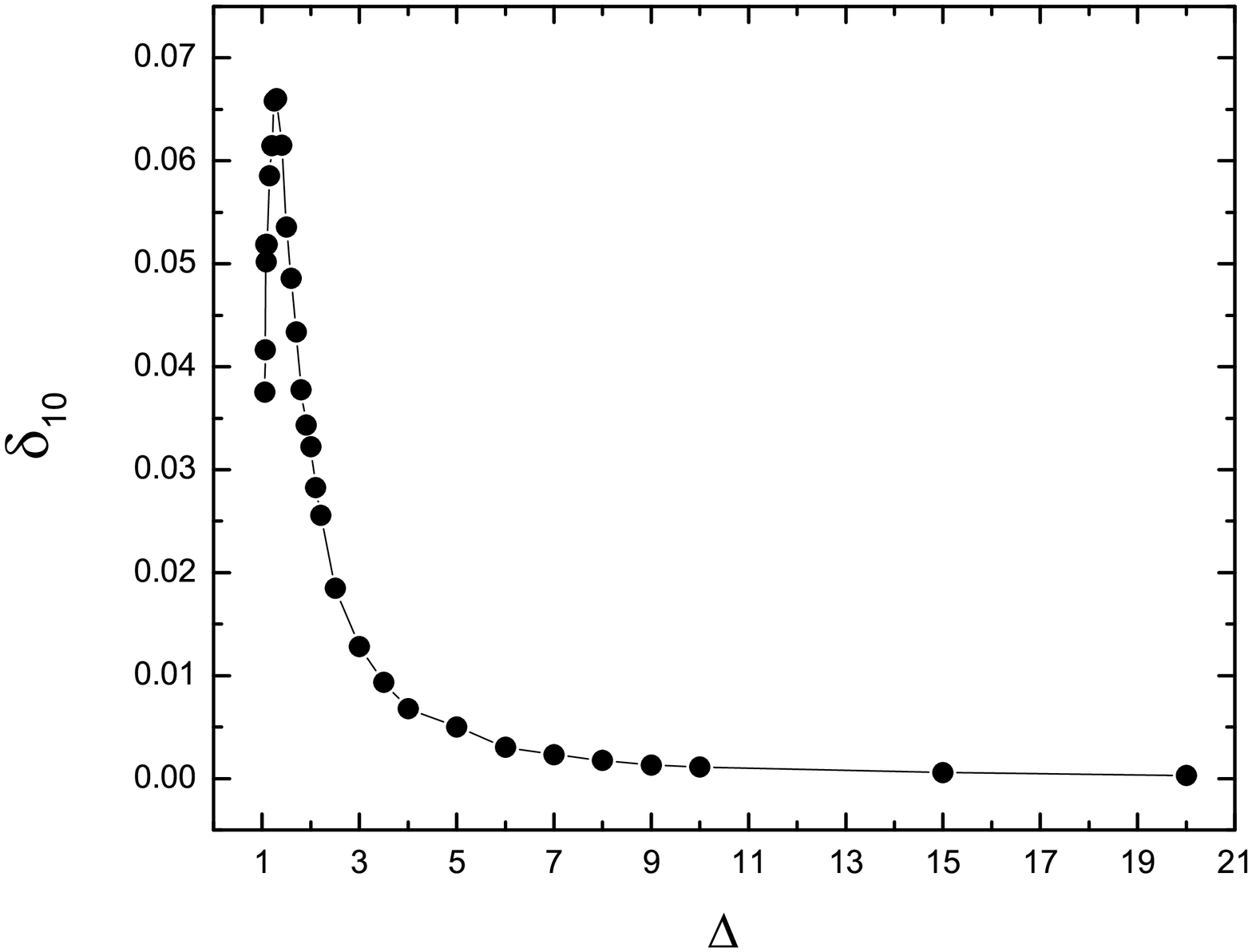}
\end{center}
\caption{$\protect\delta_{10}\left( \Delta\right) $ as a function of $\Delta$%
. Here $t_{1}=101\protect\tau$, $t_{2}=200\protect\tau$. The total number of
spins in the spin chain is $N=20$.}
\label{figure5}
\end{figure}

It is clear that if $\left\langle S_{n}^{z}\left( t\right) \right\rangle $
is a constant in the time interval $[t_{1},t_{2}]$, then $\delta_{n}\left(
\Delta\right) =0$. In general, since the initial state is not an eigenstate
of the Hamiltonian Eq. (\ref{H1}), the magnetization of each spin will
fluctuate and $\Omega_{n}\left( t\right) \neq0$. If, after long time, the
system relaxes to a stationary state that contains a DW, the magnetization
of each spin will fluctuate around its stationary value $C_{n}$. The
fluctuations are given by $\Omega_{n}\left( t\right) $. If $\left\vert
\Omega_{n}\left( t\right) \right\vert $ is large, the difference between the
actual magnetization profile at time $t$ and the stationary profile $C_{n}$
may be large. From Eq. (\ref{delta2}), it is clear that $\delta_{n}\left(
\Delta\right) $ is a measure of the deviation of $\left\langle S_{n}
^{z}\left( t\right) \right\rangle $ from its stationary value $C_{n}$,
averaged over the time interval $[t_{1},t_{2}]$. Thus, $\delta_{n}\left(
\Delta\right) $ gives direct information about the dynamics stability of the
DW.

Figure \ref{figure4} shows the distribution $\delta_{n}\left( \Delta\right) $
for different values of $\Delta$. We only show some typical results, as in
Fig. \ref{figure2}. As expected, the distribution of $\delta_{n}\left(
\Delta\right) $ is symmetric about the centre of the spin chain ($n=10.5$).

We first consider how $\delta_{n}\left( \Delta\right) $ changes with $\Delta$
for fixed $n$. From Fig.\ref{figure4}, we conclude:

1) For the spins which are not located at the DW centre, i.e., $n\neq10,11$, 
$\delta_{n}\left( \Delta\right) $ decreases if $\Delta$ becomes larger. This
means that the quantum fluctuations of these spins become smaller if we
increase the value of $\Delta$. This is reasonable because with increasing $%
\Delta$, the initial state approaches an eigenstate of the Hamiltonian for
which $\delta_{n}\left( \Delta\right) =0$ (Ising limit).

2) For the spins at the DW centre, i.e., $n=10,11$, when $\Delta$ becomes
larger and larger, $\delta_{n}\left( \Delta\right) $ first increases and
then decreases. Qualitatively, this can be understood in the following way.
When $\Delta$ is close to $1$, the magnetization at the DW centre disappears
very fast and remains zero. However, if $\Delta>>1$, the magnetization at
the DW centre will retain its initial direction, hence the behavior of \ the
spin at the DW centre will qualitatively change as $\Delta$ moves away from
the critical point $\Delta=1$. In Fig. \ref{figure5}, we plot $%
\delta_{10}\left( \Delta\right) $ ($=\delta_{11}\left( \Delta\right) $) as a
function of $\Delta$.\ It is clear that $\delta_{10}\left( \Delta\right) $
first increases as $\Delta$ increases, reaches its maximum at $\Delta=1.3$,
and then decreases as $\Delta$ becomes larger.

Now we consider the $n$-dependence of $\delta_{n}\left( \Delta\right) $ for
fixed $\Delta$. Since $\delta_{n}\left( \Delta\right) $ is a symmetric
function of $n$, we may consider only one side of the whole chain, e.g., the
spins with $n=1,2,...,N/2$. From Fig.\ref{figure4}, according to the value
of $\Delta$, there are three different regions:

1) $1.05\leq\Delta\leq1.3$: starting from the boundary ($n=1$), $\delta
_{n}\left( \Delta\right) $ first decreases, then increases, and finally
decreases again as $n$ approaches the DW centre ($n=10$). As we discussed
already, the fluctuation of the magnetization at the DW centre is small when 
$\Delta$ is close to $1$. The spin at the boundary only interacts with one
nearest spin, so it has more freedom to fluctuate. For the others, because
of the influence of the DW structure (or boundary), the fluctuations of the
spins which are near the DW (or near the boundary) are larger compared to
those of a spin located in the middle of a polarized region. Thus $%
\delta_{n}\left( \Delta\right) $ is larger if the spin is located near the
DW or near a boundary.

2) $1.3\leq\Delta\leq5$: $\delta_{n}\left( \Delta\right) $ reaches its
maximum at the DW centre. The reason for this is that in this regime the
magnetizations of all spins retain their initial direction, therefore the
spins that are far from the centre fluctuate little.

3) $5<\Delta$: in this regime (Ising limit), the initial state is very close
to the eigenstate, and the fluctuations are small, even for the spins at the
DW.

\section{Summary}

In the presence of Ising-like anisotropy, DWs in a ferromagnetic spin $1/2$
chain are dynamically stable over extended periods of time. The profiles of
the magnetization of the DW are different from the profile in the ground
state in the subspace of total magnetization $M=0$. As the system becomes
more isotropic, approaching the quantum critical point, the width of the DW
increases as a power law, with an exponent equal to $1/2$.

\end{document}